\documentclass[lettersize,journal]{IEEEtran}
\usepackage{amssymb,amsmath}
\usepackage{cite}
\usepackage{graphicx}
\usepackage{psfrag}
\usepackage{url}
\usepackage[T1]{fontenc}
\usepackage[utf8]{inputenc}
\usepackage[absolute,overlay]{textpos}
\usepackage{gensymb}
\usepackage{cases}
\usepackage[font=footnotesize]{caption}
\usepackage[font=footnotesize]{subcaption}
\usepackage{float}
\usepackage[linesnumbered,ruled,lined]{algorithm2e}
\SetKwInOut{Training}{Train}
\SetKwInOut{Testing}{Test}
\usepackage{pgf}

\usepackage[nocomma]{optidef}

\usepackage{amsthm}

\usepackage{booktabs}
\usepackage{color,soul}

\usepackage{textcomp}
\usepackage{xcolor}
\def\BibTeX{{\rm B\kern-.05em{\sc i\kern-.025em b}\kern-.08em
    T\kern-.1667em\lower.7ex\hbox{E}\kern-.125emX}}

\usepackage{tabularx}
\usepackage{makecell}
\usepackage{multirow}

\begin{document}

\title{A Multi-Task Oriented Semantic Communication Framework for Autonomous Vehicles}
\author{
	\IEEEauthorblockN{Eslam Eldeeb, Mohammad Shehab and Hirley Alves \\
	}
	\thanks{This work is partially supported by Academy of Finland, 6G Flagship program (Grant no. 346208), and the European Commission through the Horizon Europe project Hexa-X-II (Grant Agreement no. 101095759).}
	
	\thanks{Eslam Eldeeb and Hirley Alves are with Centre for Wireless Communications (CWC), University of Oulu, Finland. Email: firstname.lastname@oulu.fi. 
 
 Mohammad Shehab is with CEMSE Division, King Abdullah University of Science and Technology
(KAUST), Thuwal 23955-6900, Saudi Arabia. Email: mohammad.shehab@kaust.edu.sa}
}
\maketitle

\begin{abstract}
Task-oriented semantic communication is an emerging technology that transmits only the relevant semantics of a message instead of the whole message to achieve a specific task. It reduces latency, compresses the data, and is more robust in low SNR scenarios. This work presents a multi-task-oriented semantic communication framework for connected and autonomous vehicles (CAVs). We propose a convolutional autoencoder (CAE) that performs the semantic encoding of the road traffic signs. These encoded images are then transmitted from one CAV to another CAV through satellite in challenging weather conditions where visibility is impaired. In addition, we propose task-oriented semantic decoders for image reconstruction and classification tasks. Simulation results show that the proposed framework outperforms the conventional schemes, such as QAM-$16$, regarding the reconstructed image's similarity and the classification's accuracy. In addition, it can save up to $89 \%$ of the bandwidth by sending fewer bits.

\end{abstract}

\section{Introduction}\label{sec:introduction}

The recent advances in intelligent communication networks lead to new challenges, such as massive connectivity, massive traffic explosion, and extremely low latency~\cite{gunduz2022transmitting}. The next generation, $6$G, shall meet a wide range of requirements and use cases. Among those is the Internet of Vehicles (IoV), a key enabler for intelligent transportation systems (ITSs) \cite{SC4IoV-XuVTM2023}. A key component for IoV is vehicle-to-everything (V2X) \cite{6GV2XNoorProcIEEE2022}, which has attracted significant research interest from academia and industry. This is evidenced by the introduction of cellular V2X in the current generation via 3GPP standardization since Release 15, e.g., 5G NR V2X. 

A particular use case that will benefit substantially from such developments is connected and autonomous vehicles (CAVs) \cite{6GV2XNoorProcIEEE2022}. Typically, CAVs can operate themselves, communicate with each other, and make all the needed decisions without human intervention~\cite{10013736}. Each CAV communicates with the nearby CAVs to share information, such as traffic signs and relevant road information, that is needed to take action. Despite being able to self-drive in perfect road conditions, there are ethical concerns that the CAVs still require a human presence to take the lead when necessary and to have full control~\cite{8436265}. These scenarios impose challenging demands for current C-V2X, particularly for cases with limited coverage.

$6$G is envisioned to integrate terrestrial and non-terrestrial networks, i.e., satellites and uncrewed aerial vehicles (UAVs), and provide native intelligence \cite{6GLEOEdge-LaiVTM2023, LEOUAVJiaIOTJ2021}. The first point is relevant to V2X because of the extended coverage of UAVs or satellites and increased data rates. Current V2X standards only use satellites for localization purposes. Therefore, we expect an increase in interest in using Low-Earth Orbit (LEO) satellites for V2X use cases due to the exponential growth of such solutions in recent years \cite{LEOUAVJiaIOTJ2021, 6GV2XNoorProcIEEE2022, 6GLEOEdge-LaiVTM2023}. The second point is vital for V2X because certain scenarios are challenging to model accurately. Hence, data-driven solutions can assist in making inferences and predictions, e.g., regarding channel dynamics, network traffic, and security threats, leading to better resource allocation and network operation.
\begin{figure*}[t!]
    \centering   
    \subfloat[\label{Finland}]{\includegraphics[width=0.33\textwidth,trim={0cm 0cm 8cm 0cm},clip]{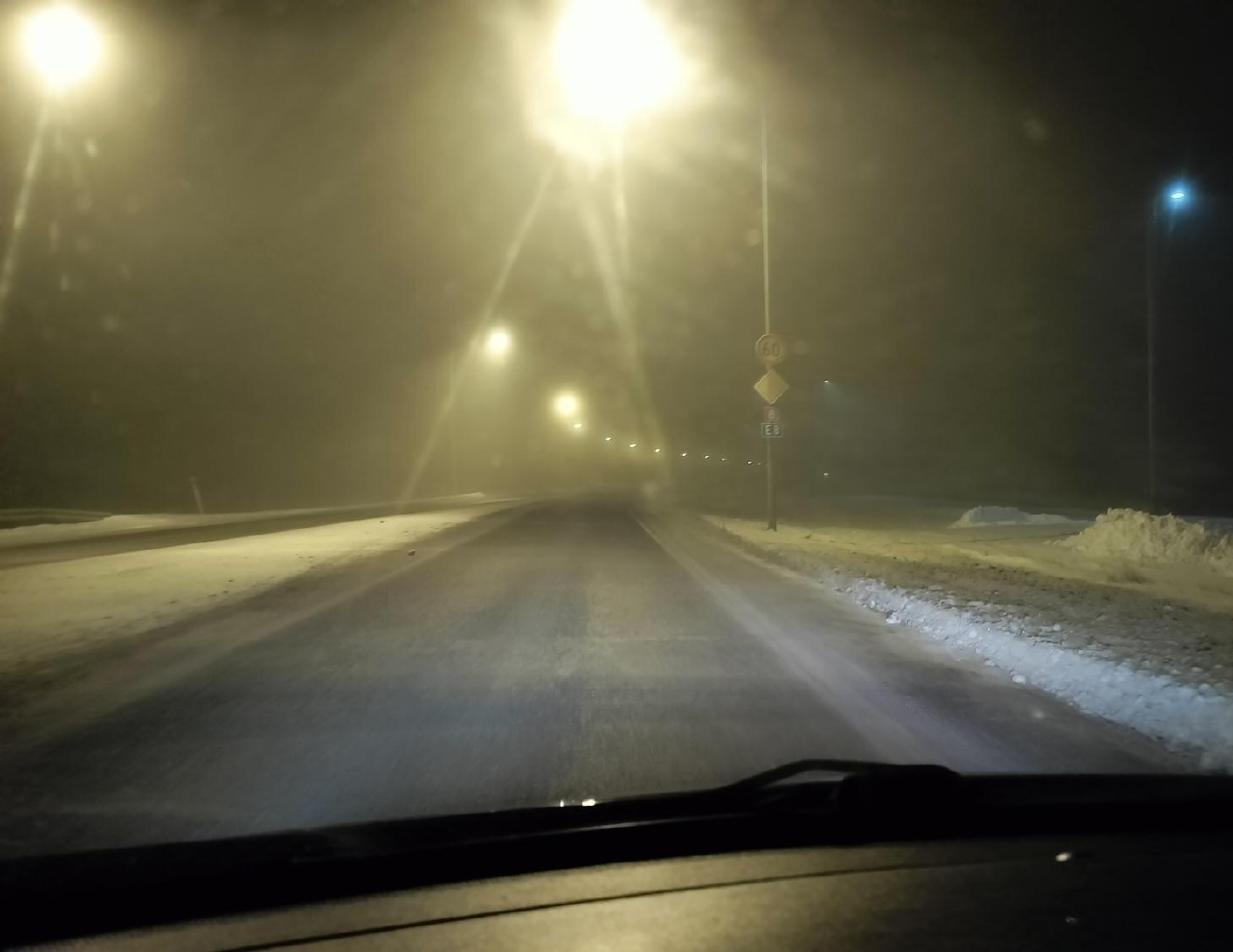}}
    \hskip -1.065ex \hspace{5mm}
     \subfloat[\label{bsys_model}]{\includegraphics[width=1.19\columnwidth,height=5.5cm,trim={0 0 0 0},clip]{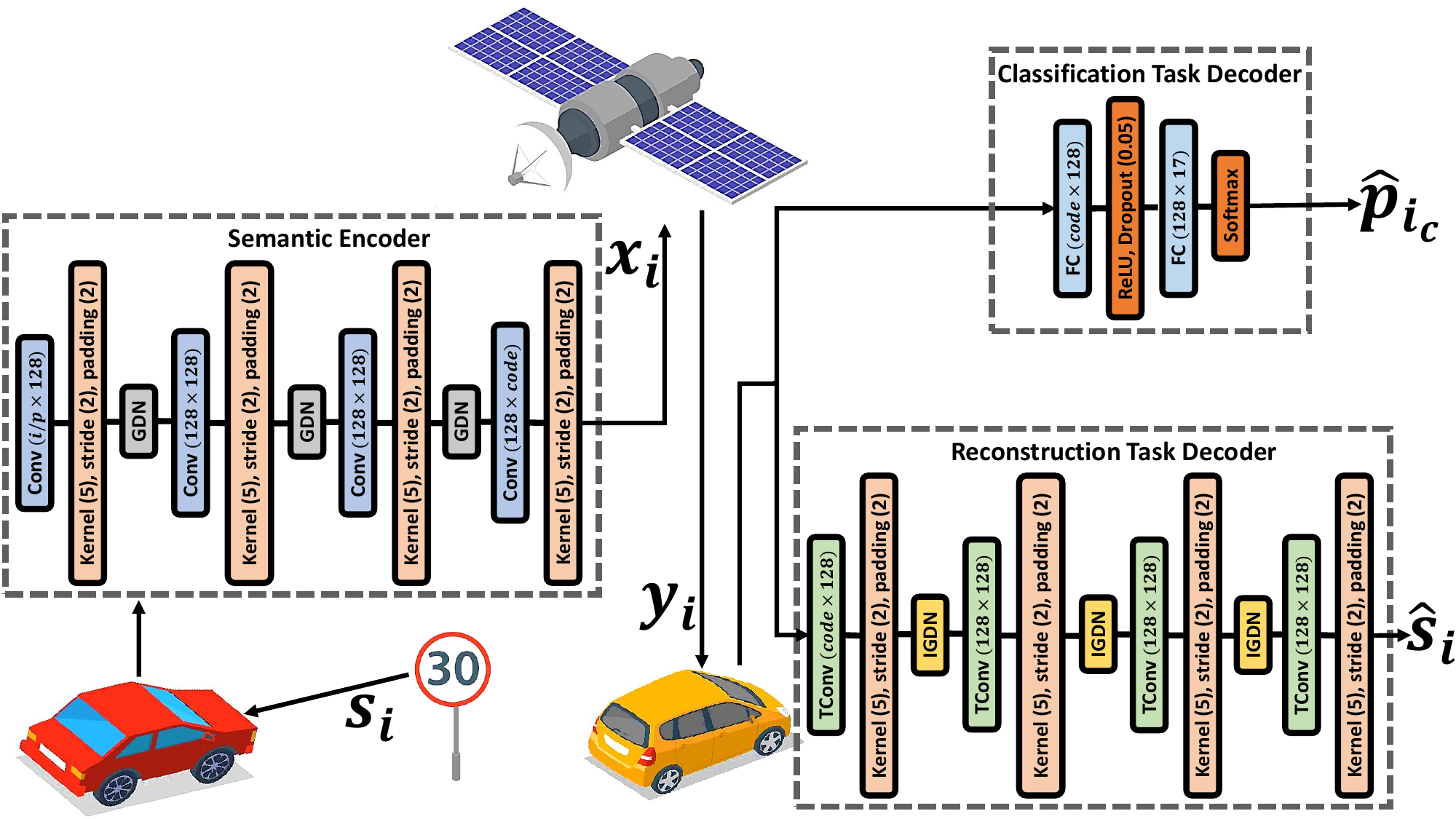} }
    
    \caption{(a) Road sign with impaired visibility, (b) The proposed multi-task semantic communication framework.} 
    \label{sys_model}
\end{figure*}

Deep neural networks are affecting system design and optimisation, prompting a transition from a model-driven approach to a data-driven approach. For instance, Yu \textit{et. al.} discusses the role of deep learning in future wireless systems and indicates that it will be critical in identifying solutions with reduced computational complexity, for instance in  optimising reflective surfaces, and multiuser beamforming \cite{RoleDLYuBITS2022}. Some even envision deep learning-based communication systems will eventually replace the conventional model-based communication systems \cite{RoleDLYuBITS2022}. Within the context of deep learning, semantic and task-oriented communication emerged aiming to extract the meaning from exchanged information  \cite{gunduz2022transmitting, zhang2022goaloriented,10049005,zhang2022wireless}. Intelligent semantic and task-oriented communication only transmits the relevant information (the semantics of the message) to the receiver, reducing the traffic and the required resources, thereby improving communication efficiency~\cite{gunduz2022transmitting}. The authors in \cite{zhang2022goaloriented} provide insights into the key differences between task/goal-oriented and semantic communications on IoT applications. In contrast, the authors in \cite{TOCShiWC2023} introduce a task-oriented communication framework for end-to-end information compression, encoding and transmission that efficiently extracts the relevant information based on assigned tasks, thus reducing the communication overhead and latency. The authors in \cite{TOCIBShaoJSAC2022} jointly optimize feature extraction, source coding, and channel coding resort and resort to information bottleneck to characterize the rate-distortion trade-off between the novelty of the encoded feature and the inference performance. They show improved rate-distortion trade-off with reduced latency, particularly relevant in dynamic channel conditions. However, information bottleneck is a computationally costly solution, requiring variational approaches to estimate the underlying joint or conditional distributions. The security aspects in semantic image transmission are discussed in~\cite{zhang2022wireless}. 

As discussed earlier, CAVs operate mostly without human intervention and collect and transmit larger volumes of data. In many cases, the raw data may not even be used by the receiver, thus generating excessive data and inefficient use of resources (e.g., spectrum and energy). Recent works have started advocating for TSC for the IoV \cite{SC4IoV-XuVTM2023, SCV2V-SuTVT2023, 10049005} to alleviate this issue. For instance, the authors \cite{SC4IoV-XuVTM2023} discuss the cooperative architecture to enable semantic communication in IoV while reducing the data traffic. In \cite{SCV2V-SuTVT2023}, authors propose a robust resource management solution for reliably exchanging semantic information in an IoV setting. Notably, \cite{10049005} introduces a system model where a CAV collects traffic signs, applies a semantic encoding and transmits it to vehicular infrastructure. The infrastructure will process the semantic encoded data and return an action. The authors proposed a convolution autoencoder (CAE) architecture as a semantic transceiver and formulated a proximal policy optimization (PPO) algorithm as a decision-maker of the received semantics. Thus, after transmitting the semantics, the receiver should make the correct decisions and reconstruct the full image of the traffic for possible human intervention. 

We propose a multi-task-oriented communication framework for a CAV network. Our contribution builds upon \cite{10049005}; however, we assume strenuous weather conditions with limited visibility of traffic signs as shown in Fig. \ref{Finland} during a snow storm in Finland. A CAV would collect traffic signs during times of clear visibility, extracts the image semantics, and transmits it to a LEO satellite constellation. The satellite broadcasts the received message to other CAVs passing in the same area. Our main contributions are
\begin{itemize}
    \item We propose a multi-task semantic communication framework that enables a semantically efficient encoding and transmission of road traffic signs between autonomous vehicles; the receiving vehicles' tasks are: image reconstructions or image classification. Once the task is completed, the CAV determines the action required. 

    \item We assume a vehicle-satellite-vehicle link. We propose two distinct frameworks for the tasks. For the image reconstruction task, we propose a CAE. However, we use a convolution neural network (CNN) with split learning for the classification task.

    \item Our results show that the proposed framework outperforms the conventional schemes regarding the amount of data transferred and classification accuracy. Our proposed framework is robust even in low SNR scenarios. This is particularly interesting when communicating with LEO satellites or under extreme weather conditions.  
\end{itemize}


\section{System model and Problem Formulation}\label{sec:sysmodel}

As shown in Fig.~\ref{bsys_model}, consider a self-driving vehicle in an urban area and a road sign $s$ being captured by this vehicle. The set of the traffic sign images is $\mathcal{S} = \{s_1, s_2, ..., s_S\}$, where $S = |\mathcal{S}|$ is the total number of images. Each traffic sign image represent a class $c$ of a set of possible classes $\mathcal{C} = \{c_1, c_2, ..., c_C\}$, where $C = |\mathcal{C}|$. Assume a given sign, $s_i$, then the vehicle extracts its semantic information, $x_i$, and transmits it to an LEO satellite\footnote{Without loss of generality, we assume that the CAV connects directly to an LEO satellite constellation and that there is always at least one satellite visible \cite{6GV2XNoorProcIEEE2022, LEOUAVJiaIOTJ2021}.}. The satellite retransmits the semantic information to another vehicle driving at the same place but at a different time; thus, the received message is denoted as $y_i$, as illustrated \figurename~\ref{sys_model}. 
The path loss between the vehicle and the satellite can be modelled using the Friss equation 
$P_r = P_t G_T G_r \frac{\lambda^2}{(4\pi d)^2}$,
where $\lambda$ is the wavelength, $P_t$ is the transmit power, $G_T$, and $G_R$ are the transmit and receive antenna gains, respectively. Then, the distance between the satellite and the vehicle is 
$d=R\left[\sqrt{\left(\tfrac{H+R}{H}\right)^2-\cos^2 \alpha}- \sin \alpha\right]$,
where $R$ is the earth's radius, $H$ is the satellite orbit height, and $\alpha$ is the satellite elevation angle \cite{Asad_MTC}. 
%

\noindent\textbf{Problem formulation:} The destination's goal is to use the received semantic information to estimate the true class of the image with the highest probability $\hat{p_c}$ or to reconstruct the image with high quality for possible human intervention. In other words, the goal is to maximize the probability of classifying the images with their true classes, i.e., minimize the error between the probability of the true class and the estimated probability of that class, and to minimize the difference between the pixels of the true image and the reconstructed images. 

We formulate the optimization problem as follows
\begin{subequations}\label{P1}
	\begin{alignat}{2}
	\mathbf{P1:}\: \: \: \: &\underset{x}{\min}   &\ &  \sum_{i=1}^{S} 1\!-\!\hat{p}_{i_c} \!+\! \!\sum_{j=1}^{J} \!\sum_{k=1}^{K} \!\left(\lvert s_i\left(j,k\right)\!-\!\hat{s}_i\left(j,k\right)\rvert \!\right), \label{P1:a}
	\ \\
	&\text{s.t.}   &      & 0 \leq \hat{p}_{i_c} \leq 1,\label{P1:b}\\
		& & & 0 \leq s_i\left(j,k\right), \hat{s}_i\left(j,k\right) \leq 255, \label{P1:c}\\
		& & & \lvert s_i\rvert = \lvert \hat{s}_i\rvert, \label{P1:d} 
	\end{alignat}
\end{subequations}
where $i$ is the $\text{i}^\text{th}$ image, $c$ corresponds to the $\text{c}^\text{th}$ class, $\hat{p}_{ic}$ is the predicted probability of the true class $c$ of an image $i$, $J$ is the horizontal size of the image, and $K$ is the vertical size of the image. We are converting the image to a grey scale; therefore, there are only horizontal and vertical axes for the images (without including the depth). Then, constraint \eqref{P1:b} ensures that the predicted probability ranges between $0$ and $1$, \eqref{P1:c} establishes the fact that pixels ranges are between $0$, which represents the black colour and $255$, which represents white colour, and \eqref{P1:d} confirms that the original and reconstructed images have the same sizes.

\section{The Multi-Task Semantic Framework}\label{sec:semantic} 

In this section, we present the proposed semantic encoder and the task-oriented semantic decoders.

\noindent\textbf{Semantic Encoder:} The semantic encoder is a shared layer where the images are encoded and transmitted to the channel, as in Fig.~\ref{sys_model}. It takes the traffic sign image $s_i$ as its input and outputs a semantic representation to be transmitted as follows
\begin{align}
\label{ENCODER_EQUATION}
    x_i = f_e\left(w_e s_i + b_e\right),
\end{align}
where $w_e$ represents the encoder weights, $b_e$ is the bias, and $f_e$ is a non-linear function that maps the traffic sign image $s_i$ to the semantic encoded message $x_i$. Herein, we use $f_e$ as the encoder part of a convolution autoencoder consisting of convolution layers with kernels and stride layers. In addition, we use generalized normalization transformations (GDNs)~\cite{balle2016density} as normalization blocks.

\noindent\textbf{Task-Oriented Decoder:} The decoder consists of multiple task-oriented semantic decoders. In particular, it receives the semantic message 
$y_i = h x_i + n$,
$h$ is the channel coefficient and follows Rayleigh distribution, and $n$ is zero mean Gaussian noise. 
For the reconstruction task, the decoder reconstructs the traffic sign image $\hat{s_i}$ from $y_i$ as 
\begin{align}
\label{DECODER_EQ_1}
    \hat{s_i} = f_r\left(w_r y_i + b_r\right),
\end{align}
where $w_r$ represents the reconstruction decoder weights, $b_r$ is the bias, and $f_r$ is a non-linear function that maps the received message $y_i$ to the reconstructed traffic sign $\hat{s_i}$. Herein, we use $f_r$ as the decoder part of a convolution autoencoder, consisting of transpose convolution layers with kernels, stride layers and inverse GDN (IGDN). The mean square error (MSE) is applied to update the weights of the reconstruction task as
\begin{align}\label{MSE_LOSS}
    L_r(w_r) = \sum_i \left( s_i - \hat{s}_i \right)^2.
\end{align}

For the classification task, the decoder classifies the traffic sign from $y_i$ by calculating the class probabilities as 
\begin{align}
\label{DECODER_EQ_2}
    \hat{p}_{i_c} = f_c\left(w_c y_i + b_c\right),
\end{align}
where $w_c$ represents the classification decoder weights, $b_c$ is the bias, and $f_c$ is a non-linear function that maps the received message $y_i$ to the estimated traffic sign class probability $\hat{p}_{i_c}$. Herein, we use $f_c$ as a fully connected neural network with a softmax function to produce the class probabilities. We use the cross-entropy loss function to update the weights of the classification task
\begin{align}\label{CE_LOSS}
    L_c(w_c) = - \sum_i \sum_c p_{i_c} \: \log\left(\hat{p}_{i_c}\right),
\end{align}
where $p_{i_c}$ is the $c$ element in the one-hot encoded true class vector and $\hat{p}_{i_c}$ is the predicted probabilities of that class. \textbf{Algorithm~\ref{alg1}} summarizes the proposed multi-task semantic transmission scheme.

\begin{algorithm}[!t]
\SetAlgoLined

\KwData{Traffic signs dataset $S$.}

Initialize model parameters $w_e,b_e,w_r,b_r,w_c,b_c$.


\For{epochs = $1$,...,$50$}{

    Feedforward the input $s_i$ through the semantic encoder $f_e$ using~\eqref{ENCODER_EQUATION}.
    
    Feedforward the received message $y_i$ through the semantic decoders $f_r$ and $f_c$ using~\eqref{DECODER_EQ_1} and~\eqref{DECODER_EQ_2}.

    Backpropagate through the network.

    \eIf{\text{task} \textbf{is} \text{classification}}{
        compute the cross entropy loss using~\eqref{CE_LOSS}.
    }
    {
        Compute the MSE using~\eqref{MSE_LOSS}.
    }

    Compute the gradient

    Update the model parameters in step $1$ using Adam optimizer
}




\caption{The multi-task semantic communication algorithm.}
\label{alg1}  
\end{algorithm}

\noindent\textbf{Key Performance Indicators:} To measure the performance of the proposed model compared to the baseline models, we introduce performance metrics for classification and regression. These metrics include the classification accuracy, confusion matrix, precision, recall, and $f1$-score~\cite{9504554}. On the other hand, we utilize the structural similarity index measure (SSIM) to evaluate the image reconstruction as it measures the similarity between two images~\cite{1284395}. The SSIM is calculated as
\begin{align}
\label{SSIM_INDEX}
    \mathrm{SSIM}(s,\hat{s}) = \frac{\left( 2 \mu_s \mu_{\hat{s}} + c_1\right) \left( 2 \sigma_{s \hat{s}} + c_2\right)}{\left( \mu^2_s + \mu^2_{\hat{s}} + c_1 \right) \left( \sigma^2_s + \sigma^2_{\hat{s}} + c_2 \right)},
\end{align}
where $\mu_s$ is the mean of $s$, $\mu_{\hat{s}}$ is the mean of $\hat{s}$, $\sigma^2_s$ is the variance of $s$, $\sigma^2_{\hat{s}}$, and $c_1$ and $c_2$ are two variables that stabilize the division. For instance, $L$ is the dynamic range of the pixels, and it is calculated as $L = 2$ in the case of gray-scale images and $L = 2^3$ in RGB images. In addition, $c_1 = \left( k_1 L \right)^2$ and $c_2 = \left( k_2 L \right)^2$, where $k_1$ and $k_2$ are fixed numbers typically set to $0.01$ and $0.03$, respectively.

\section{Numerical Results and Discussion}\label{sec:results}

This section presents the simulation results of the proposed multi-task semantic model compared to different baseline models. 
%
%
Herein, we use the term $\text{code} = l$ to describe the proposed semantic framework, where $l$ is the size of the last convolution layer in the encoder. The baseline model adopts Quadrature Amplitude Modulation $16$ $\left(\text{QAM}~16\right)$ for the image reconstruction task. On the other hand, we adopt support vector machine $\left(\text{QAM}~16 + \text{SVM}\right)$ and neural network $\left(\text{QAM}~16 + \text{NN}\right)$ as baselines classifiers for the image classification task. The simulations are performed via Pytorch framework on NVIDIA Tesla V100 GPU using the Chinese Traffic Sign Database~\cite{TRAFFIC_SIGN_DATA}, and $17$ traffic signs are selected from the whole dataset for illustration. Table~\ref{tab:Sem} presents the simulation parameters.

\begin{table}[t!]
\centering
\caption{The simulation parameters.}
\label{tab:Sem}
\begin{tabular}{cc|cc}
\toprule
\textbf{Parameter}                                    & \textbf{Value} & \textbf{Parameter}                                    & \textbf{Value}\\ \midrule
$R$ & $6378$ Km & $H$ & $780$ Km \\
$G_T$ & $10 \: \text{dB}_\text{i}$ & $G_r$ & $0 \: \text{dB}_\text{i}$ \\

$\alpha$ & $75\degree$ & Carrier frequency $f_c$ & $868$ MHz \\
Train size & $1800$ & Test size & $134$ \\

Validation size & $90$ & Input size & $34 \times 34$ \\

Encoder & $4$ Conv. layers  & $\#$ neurons & $128$ \\

Kernel size & $5$ & stride & $2$ \\

Padding & $2$ & Reconst. decoder & $4$ Conv. Transpose \\

Class. decoder & $2$ FC layers & $\#$ neurons & $128$ \\

Batch & $32$ & Epochs & $50$ \\

Optimizer & Adam & Learning rate & $0.001$ \\

Class. loss & MSE & Reconst. loss & CrossEntropy\\

\bottomrule
\end{tabular} 
\end{table}


\newcommand{\mytab}{
\footnotesize
  \begin{tabular}{|l|l|l|l|}
\Xhline{2\arrayrulewidth}
 & Prec. & Rec. & f1-Sc.\\ 
\Xhline{2\arrayrulewidth}
$0$ & $1.00$ & $1.00$ & $1.00$\\
\hline                              

$1$ & $1.00$ & $1.00$ & $1.00$\\
\hline                              

$2$ & $0.88$ & $0.88$ & $0.88$\\
\hline                              

$3$ & $1.00$ & $1.00$ & $1.00$\\
\hline                              

$4$ & $0.89$ & $1.00$ & $0.94$\\
\hline                              

$5$ & $0.91$ & $0.83$ & $0.87$\\
\hline                              

$6$ & $1.00$ & $1.00$ & $1.00$\\
\hline                              

$7$ & $0.91$ & $0.91$ & $0.91$\\
\hline                              

$8$ & $1.00$ & $1.00$ & $1.00$\\
\hline                       

$9$ & $1.00$ & $1.00$ & $1.00$\\
\hline                              

$10$ & $0.93$ & $0.93$ & $0.93$\\
\hline                              

$11$ & $1.00$ & $1.00$ & $1.00$\\
\hline

$12$ & $1.00$ & $0.86$ & $0.92$\\
\hline

$13$ & $1.00$ & $1.00$ & $1.00$\\
\hline

$14$ & $1.00$ & $1.00$ & $1.00$\\
\hline

$15$ & $1.00$ & $1.00$ & $1.00$\\
\hline

$16$ & $0.90$ & $0.91$ & $0.95$\\
\hline
\end{tabular} \vspace{-0mm}
}   

\begin{figure}[t!]
    \centering
    \subfloat[Classification accuracy\label{ACC_Class}]{\includegraphics[width=0.42\textwidth]{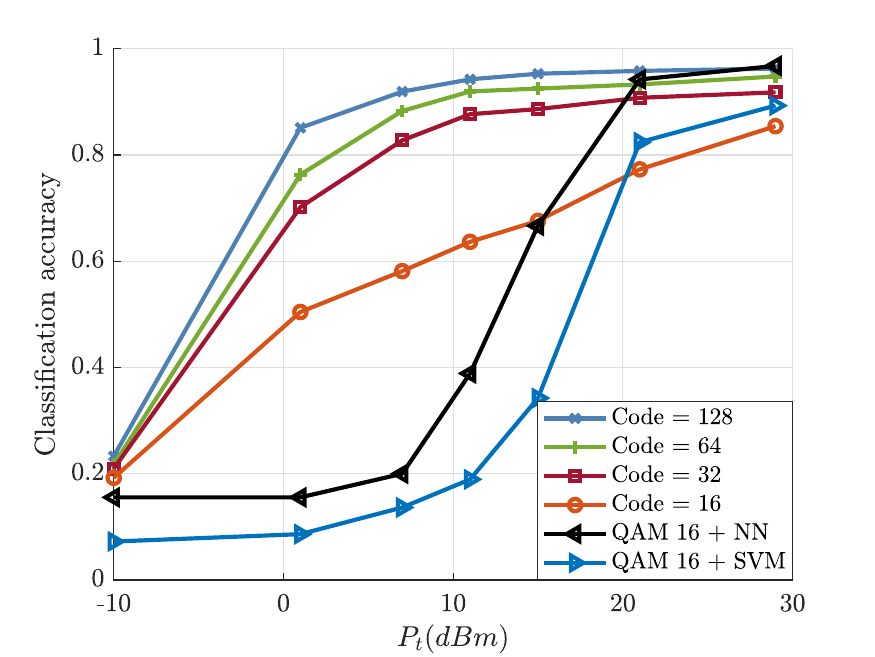}}
    \hskip -22ex
    \vskip 0.5ex
    \subfloat[Confusion matrix\label{CM_Class}]{\includegraphics[width=0.31\textwidth,trim={3cm 0 1.4cm 0},clip]{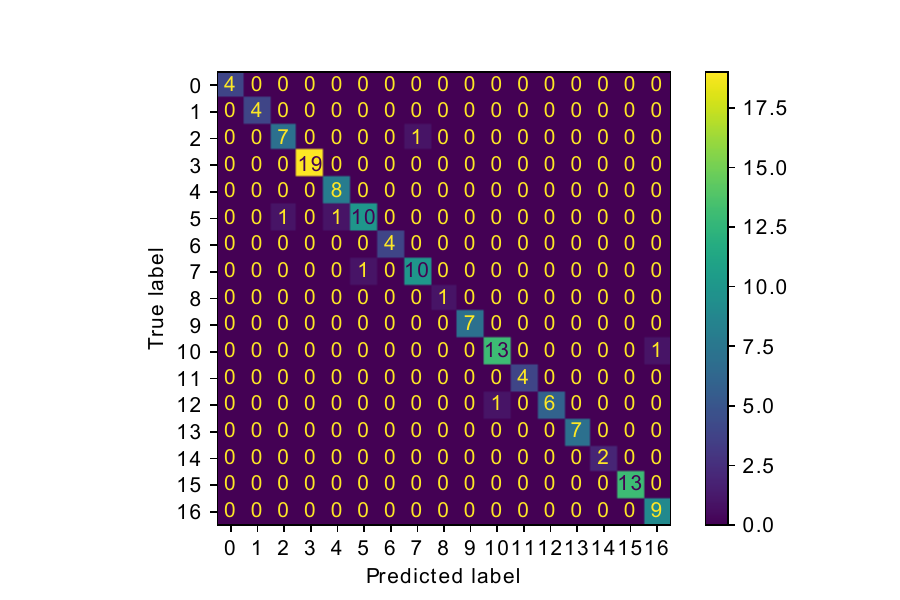}}
    \hskip -1.065ex
    \subfloat[][Classification report\label{table_class}]{\mytab}
    \hskip -1.9ex
    \caption{The classification accuracy of the proposed algorithm compared to the conventional schemes, the confusion matrix and the classification report of testing the semantic model at $P_t = 15 \: dBm$ and a codeword of $64$.}
    \label{Class_Fig} \vspace{-2mm}
\end{figure}

\subsection{Image Classification}
Fig.~\ref{Class_Fig} shows the classification performance analysis of the proposed scheme compared to the baseline models. In particular, Fig.~\ref{ACC_Class} depicts the classification accuracy of the proposed method with different codewords compared to the cases $\text{QAM}~16 + \text{SVM}$ and $\text{QAM}~16 + \text{NN}$ transmissions. Herein, the proposed model outperforms the baseline models; larger code sizes typically render better classification accuracy. Note that for low transmit power levels, which is the scenario of interest for the ground to satellite uplink and downlink, the QAM~$16$ models fail to classify the traffic signs. Meanwhile, all the schemes tend to converge in high transmission power. 

In Fig.~\ref{CM_Class}, we present the confusion matrix of the proposed model on the test data with $P_t = 15 \: dBm$ and a code of $64$ bits. It can be observed that the diagonal has the majority of the numbers, which indicates the true samples for each class. Moreover, Fig.~\ref{table_class} shows the precision, recall and $f1$-score for each traffic sign class using the proposed model on the test data with $P_t = 15 \: dBm$ and a code of $64$ bits. Interestingly, most of the precision, recall and $f1$-score value-range are pretty high $\left[0.90,1\right]$, which indicates that the model is well generalized across all classes.

\begin{figure}[t!]
    \centering
    \subfloat[SSIM\label{SSIM_Rec}]{\includegraphics[width=0.4\textwidth]{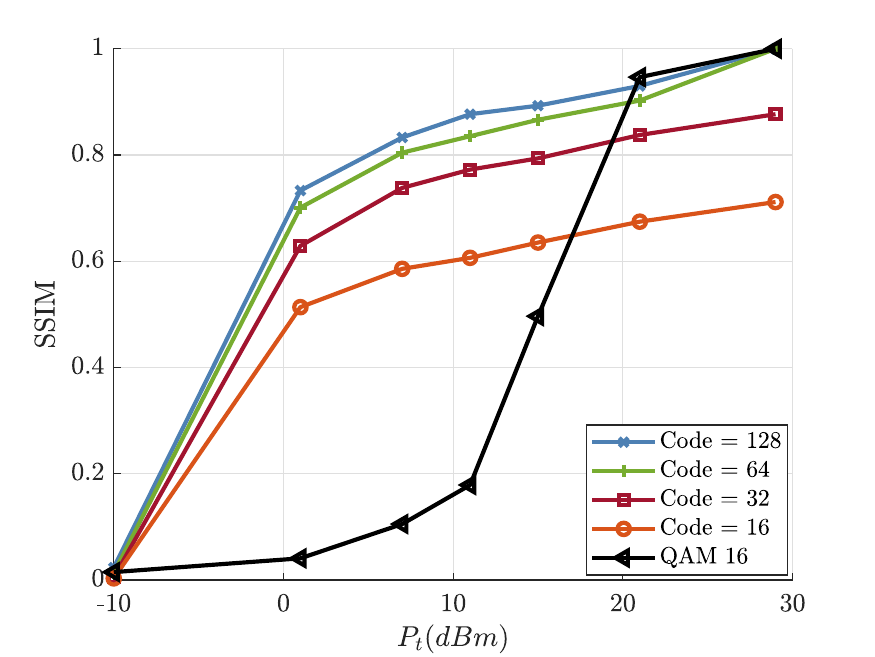}}
    \hskip -1.9ex
    \subfloat[Total transmitted byte\label{Bytes_Trans}]{\includegraphics[width=0.4\textwidth,trim={0 0 0 0},clip]{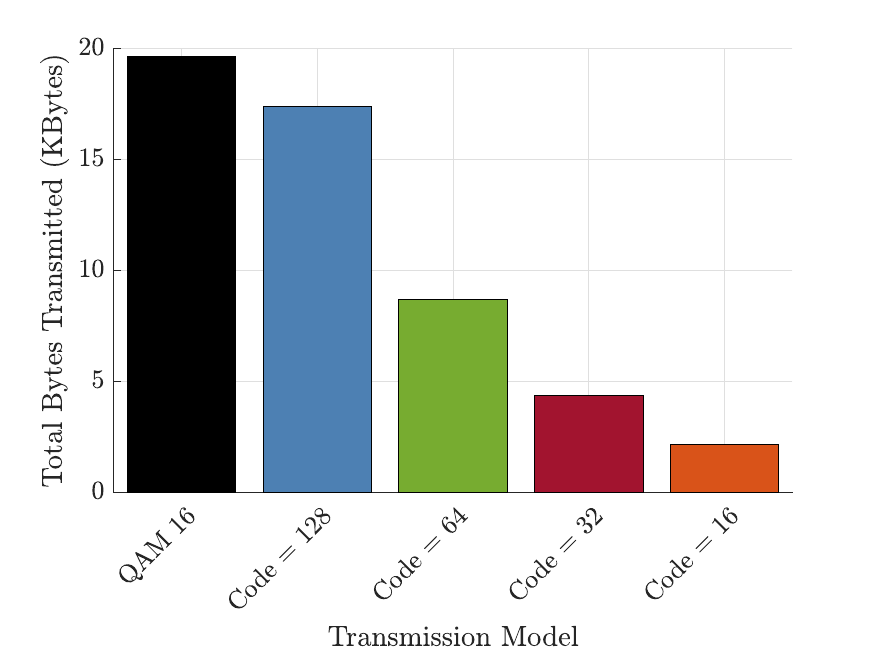}}
    \hskip -1.9ex
    \caption{The SSIM of the proposed algorithm compared to the conventional schemes and the total transmitted bytes.}
    \label{Rec_Fig} \vspace{-2mm}
\end{figure}

\subsection{Image Reconstruction}
Fig.~\ref{Rec_Fig} presents the image reconstruction performance analysis of the proposed scheme compared to $\text{QAM}~16$. In particular, Fig.~\ref{SSIM_Rec} compares the SSIM scores of the proposed method with different codewords; as the size of the code increases, the model achieves higher SSIM scores. In the satellite low SNR region of interest, the proposed model achieves higher SSIM scores than the $\text{QAM}~16$, which fails to reconstruct the images efficiently. For example, at $P_t = 1 \: dBm$, which is considered a highly noisy scenario, the proposed model with a codeword of $32$ reconstructs the images with an SSIM score of $63 \%$, whereas the $\text{QAM}~16$ achieves around $5 \%$ only. Moreover, Fig.~\ref{Bytes_Trans} illustrates the total amount of bytes transmitted using each scheme. Note that using the codewords of $128, 64, 32, 16$ saves around $11 \%, 56 \%, 78 \%, 89 \%$, respectively, of data transmission compared to the baseline $\text{QAM}~16$ scheme.

\begin{figure}[t]
    \centering
    \hskip -1.3ex \subfloat[ \: \: \: \: \: \: \:\label{Orig_1}]{\includegraphics[scale=0.235,trim={3.75cm 1.25cm 0 0},clip]{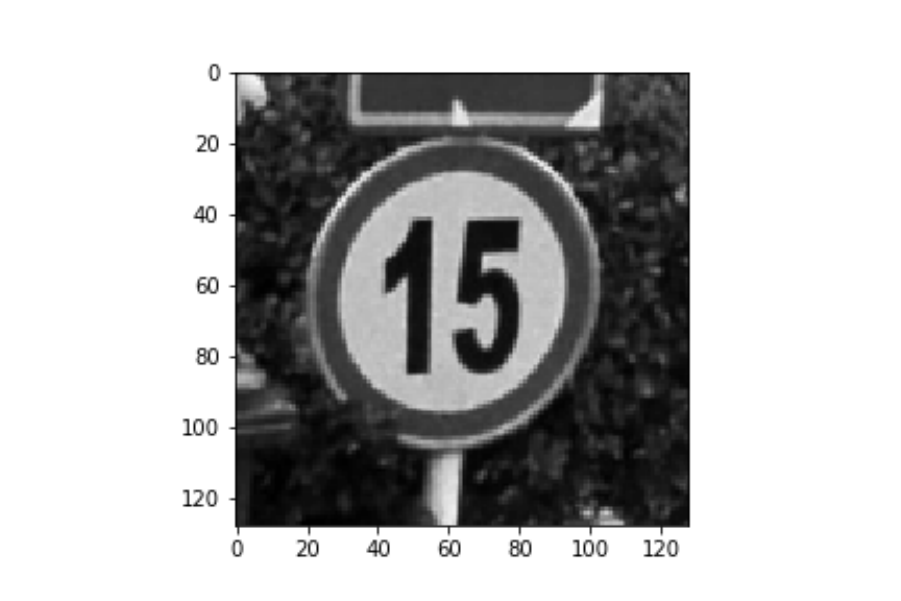}}
    \hskip -7.5ex
    \subfloat[ \: \: \: \: \: \: \:\label{Orig_2}]{\includegraphics[scale=0.235,trim={3.75cm 1.25cm 0 0},clip]{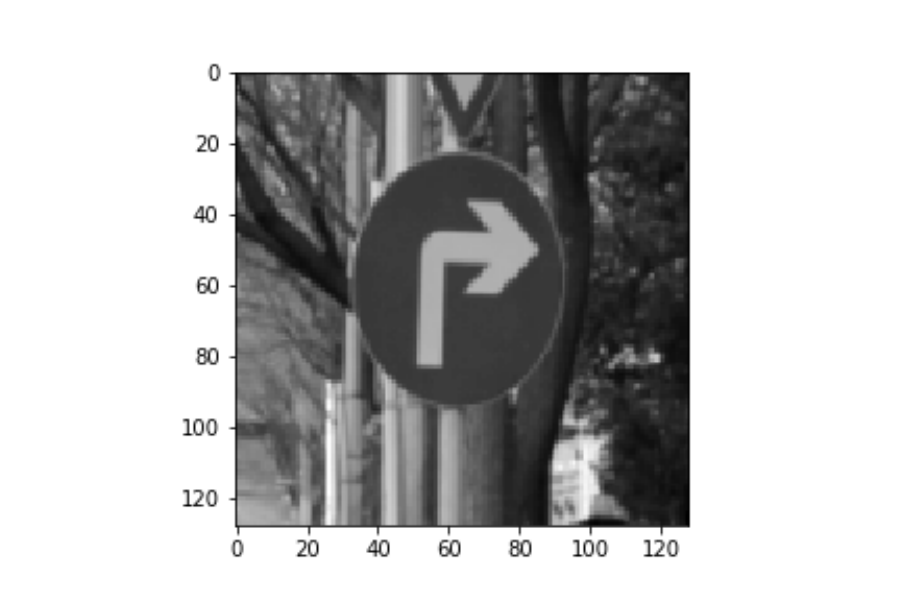}}
    \hskip -7.5ex
    \subfloat[ \: \: \: \: \: \: \:\label{Orig_3}]{\includegraphics[scale=0.235,trim={3.75cm 1.25cm 0 0},clip]{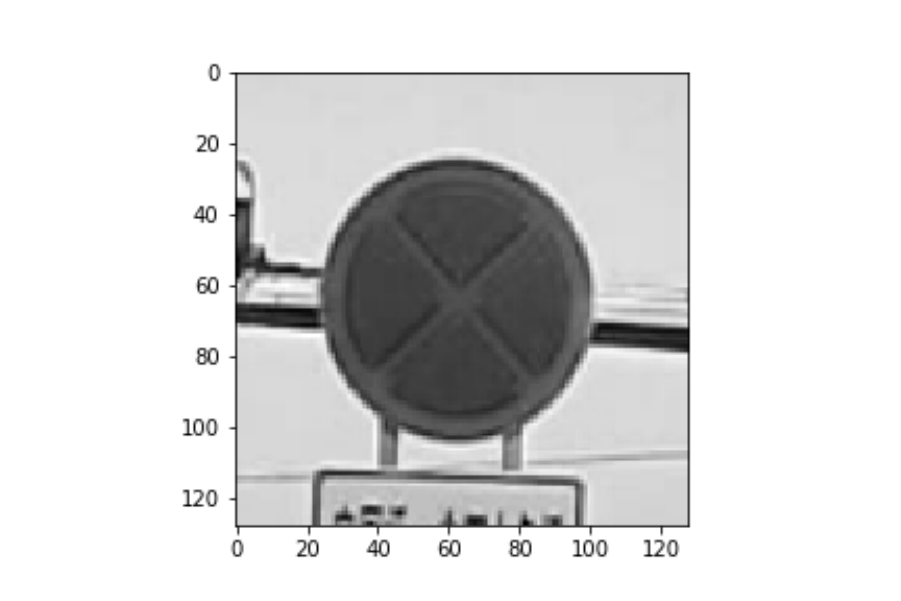}}
    \hskip -7.5ex
    \subfloat[ \: \: \: \: \: \:\label{Orig_4}]{\includegraphics[scale=0.235,trim={3.75cm 1.25cm 0 0},clip]{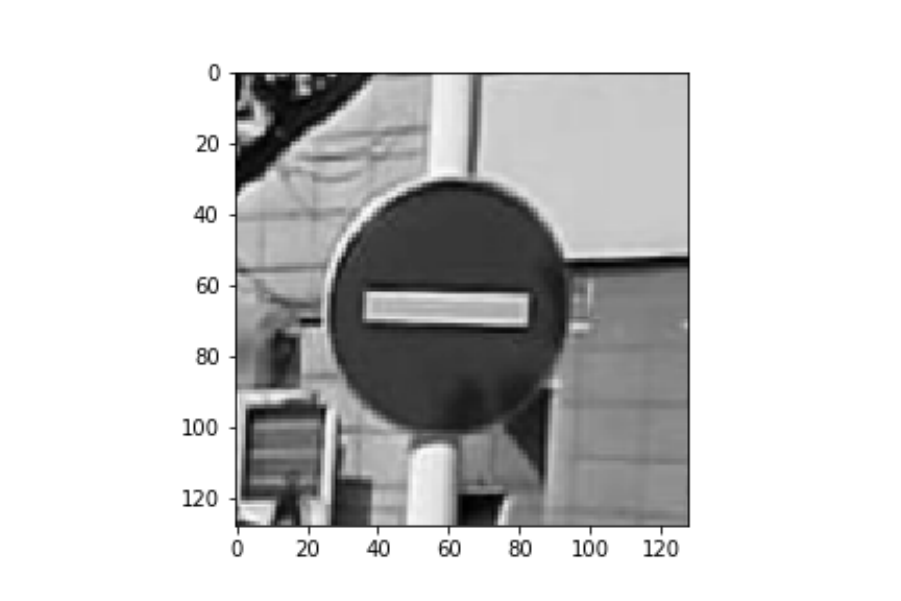}}
    \hskip -7.5ex
    \subfloat[\:\label{Orig_5}]{\includegraphics[scale=0.235,trim={3.75cm 1.25cm 4cm 0},clip]{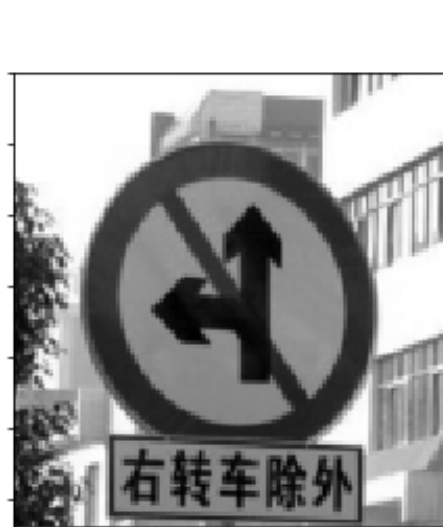}}
    
    \vskip -0.05ex
    
    \hskip -1.3ex \subfloat[ \: \: \: \: \: \: \:\label{AE_1}]{\includegraphics[scale=0.235,trim={3.75cm 1.25cm 0 0},clip]{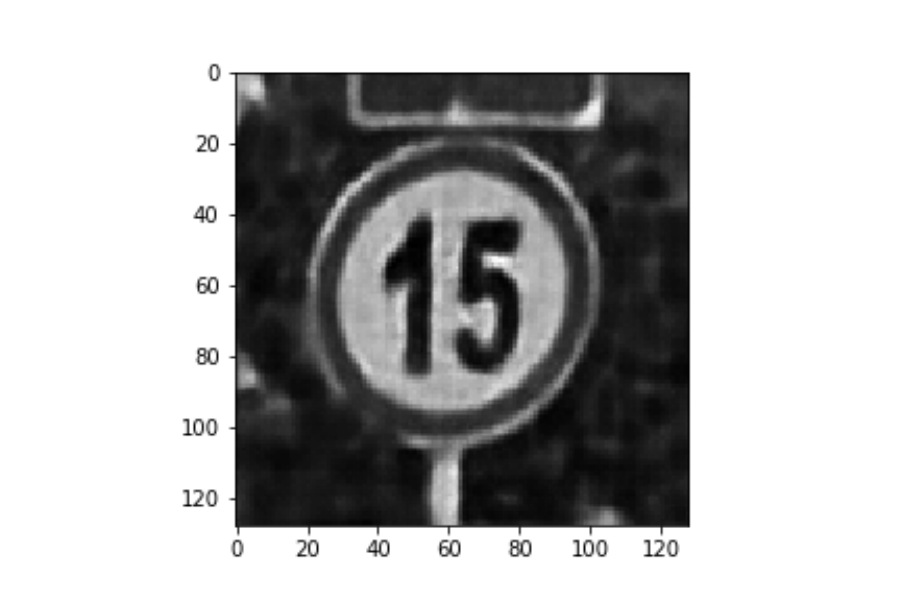}}
    \hskip -7.5ex
    \subfloat[ \: \: \: \: \: \: \:\label{AE_2}]{\includegraphics[scale=0.235,trim={3.75cm 1.25cm 0 0},clip]{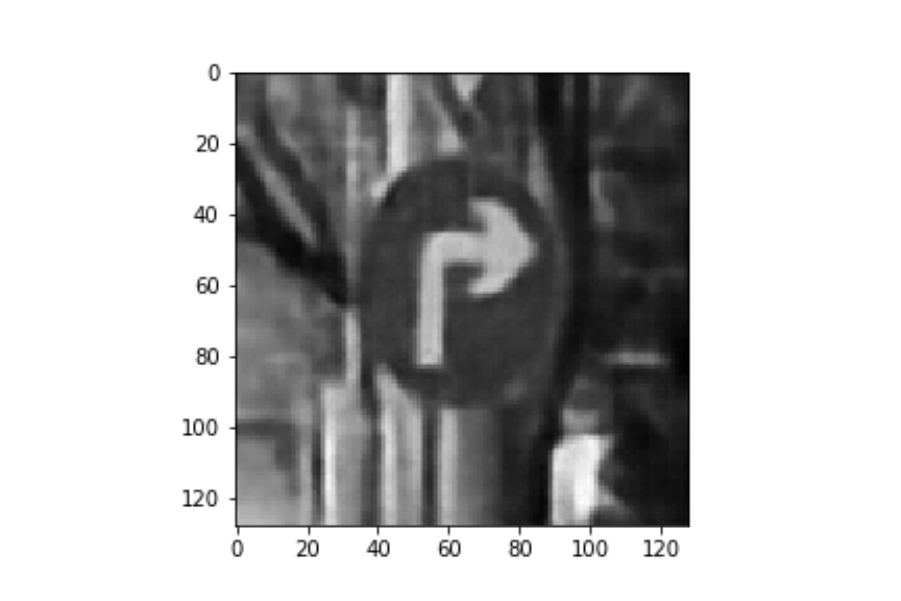}}
    \hskip -7.5ex
    \subfloat[ \: \: \: \: \: \: \:\label{AE_3}]{\includegraphics[scale=0.235,trim={3.75cm 1.25cm 0 0},clip]{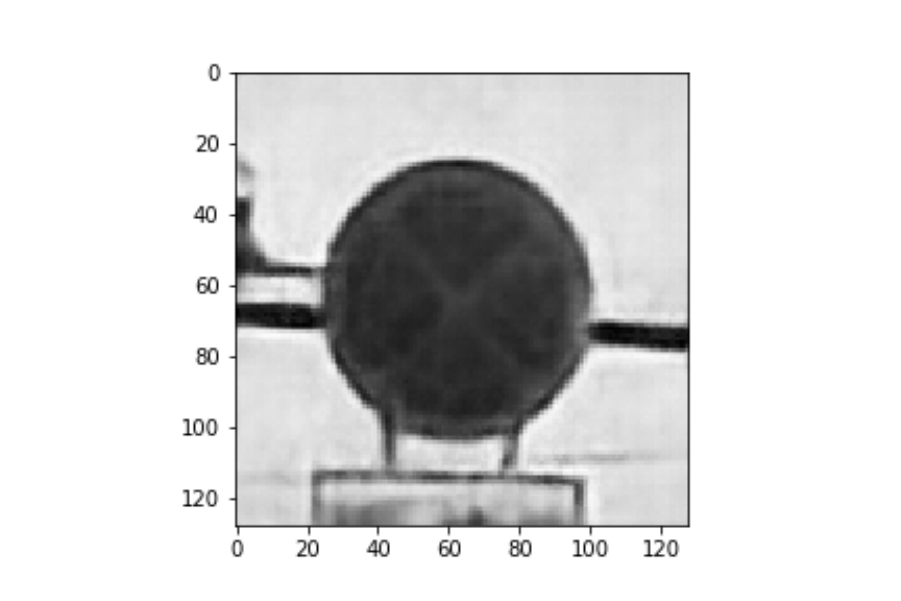}}
    \hskip -7.5ex
    \subfloat[ \: \: \: \: \: \: \:\label{AE_4}]{\includegraphics[scale=0.235,trim={3.75cm 1.25cm 0 0},clip]{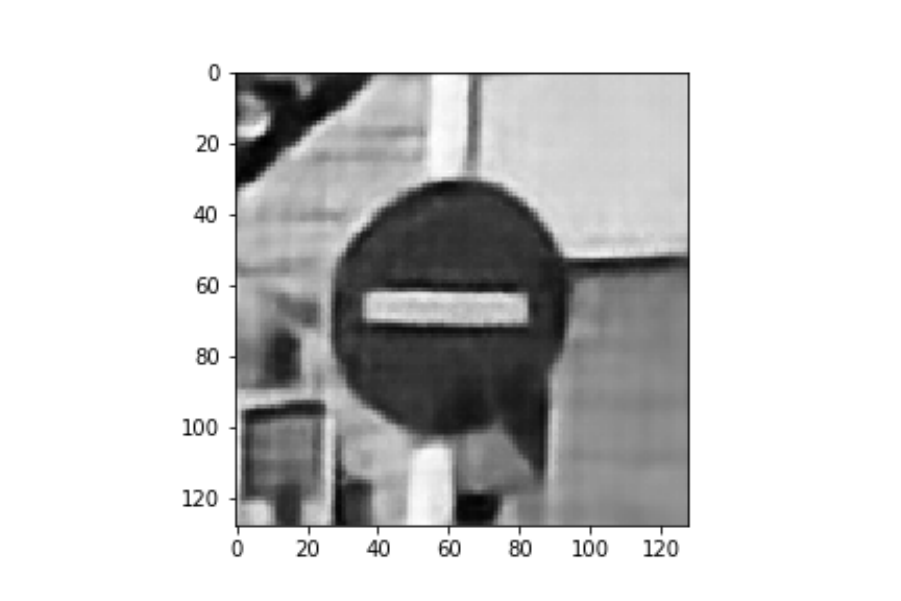}}
    \hskip -7.5ex
    \subfloat[\:\label{AE_5}]{\includegraphics[scale=0.235,trim={3.75cm 1.25cm 4cm 0},clip]{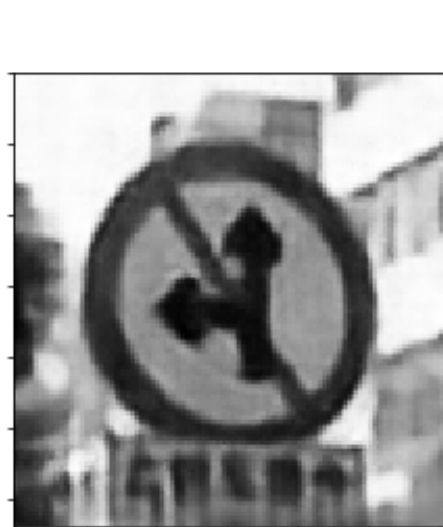}}

    \vskip -0.05ex

    \hskip 1.6ex \: \subfloat[ \: \: \: \: \: \: \: \label{QAM_1}]{\includegraphics[scale=0.235,trim={3.55cm 1.25cm 0.2cm 0},clip]{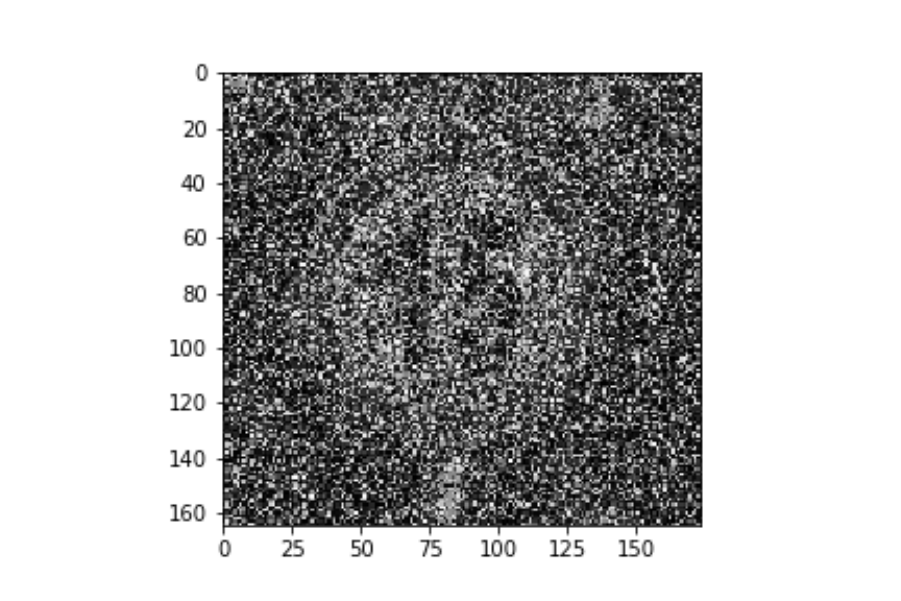}}
    \hskip -7.5ex
    \subfloat[ \: \: \: \: \: \: \:\label{QAM_2}]{\includegraphics[scale=0.235,trim={3.4cm 1.25cm 0.8cm 0},clip]{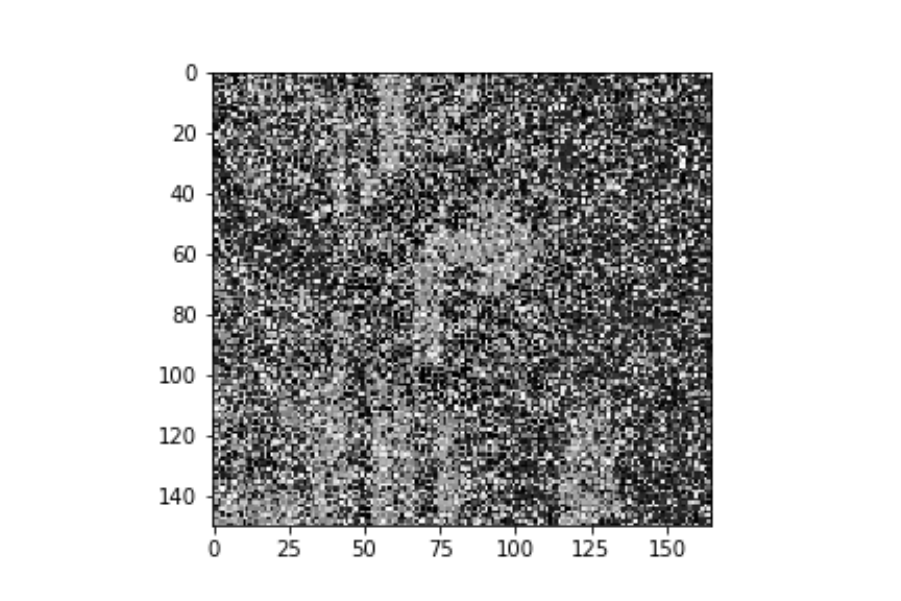}}
    \hskip -7ex
    \subfloat[ \: \: \: \: \: \: \:\label{QAM_3}]{\includegraphics[scale=0.235,trim={3.7cm 1.25cm 0 0},clip]{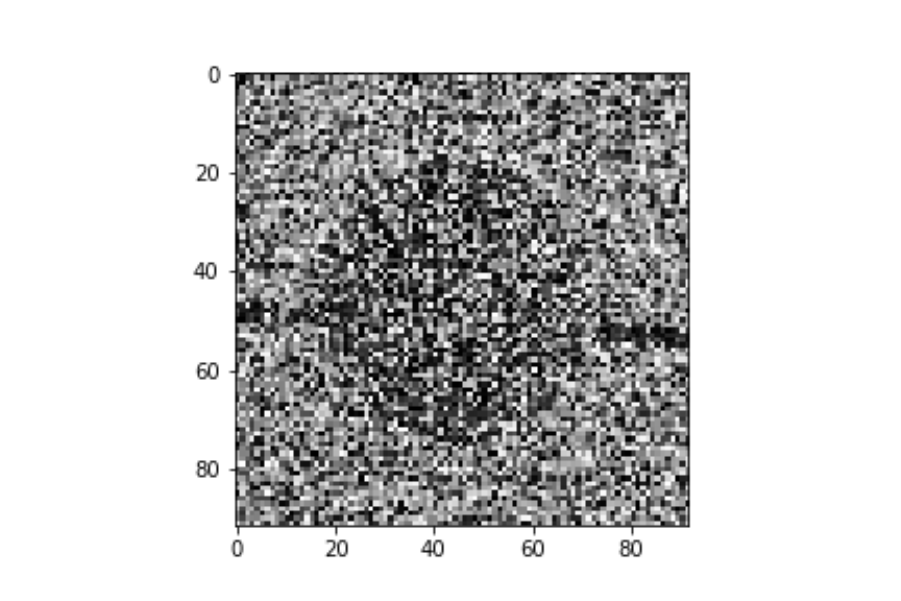}}
    \hskip -7.5ex
    \subfloat[ \: \: \: \: \:\label{QAM_4}]{\includegraphics[scale=0.235,trim={2.25cm 1.25cm 0 0},clip]{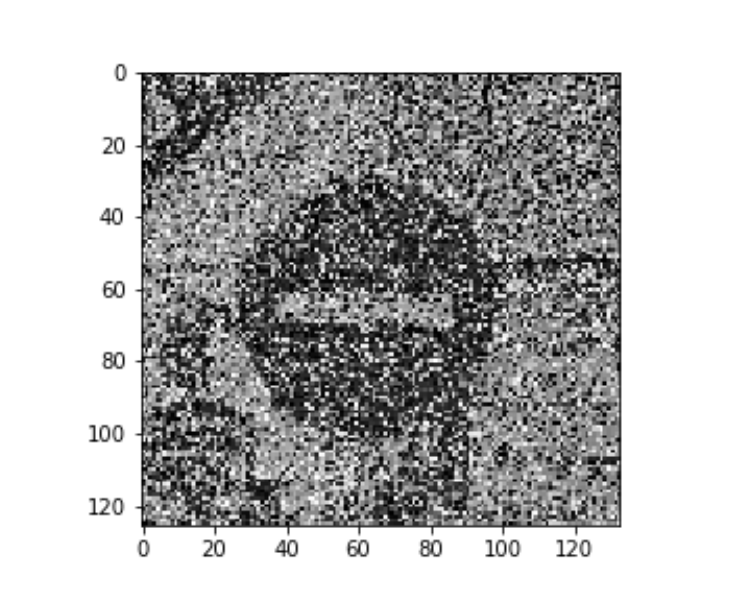}}
    \hskip -7.12ex
    \: \subfloat[ \: \: \: \:\label{QAM_5}]{\includegraphics[scale=0.245,trim={2.95cm 1.25cm 0 0},clip]{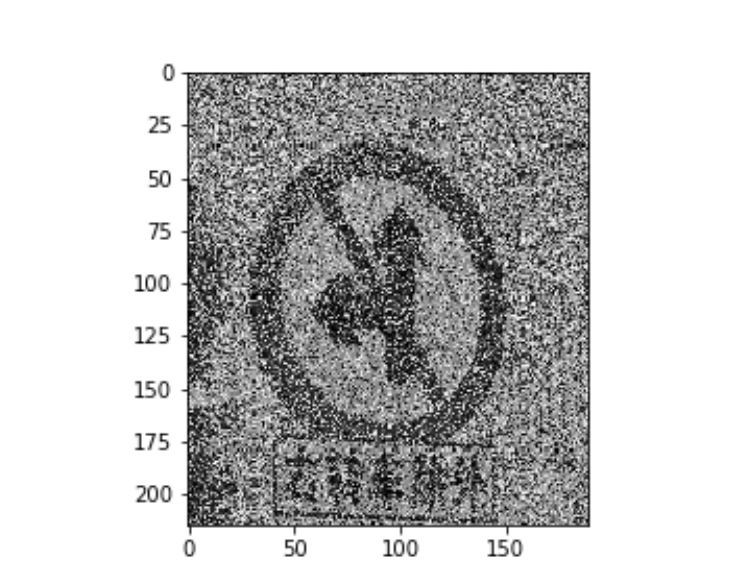}}
    
    \caption{Some examples of reconstructed images using different transmission schemes. The first row shows the original images. The second row shows the reconstructed images using the proposed model at $P_t = 15 \: dBm$ and a codeword of $64$. The bottom row shows the reconstructed images using JPEG+QAM 16.}
    \label{Reconstruct_Images} 
\end{figure}

Fig.~\ref{Reconstruct_Images} depicts some visual examples of reconstructed images using the proposed model at $P_t = 15 \: dBm$ and a codeword of $64$ compared to the $\text{QAM}~16$ scheme. The reconstructed images using $\text{QAM}~16$ are very noisy compared to the proposed scheme. The proposed scheme reconstructs the images with acceptable quality.

\section{Conclusions}\label{sec:conclusions} 

This letter presented a multi-task semantic communication framework that transmits traffic sign images through an autonomous vehicle-satellite-vehicle link. The receiver performs image reconstruction and classification tasks on the received message. Simulation results demonstrated that the proposed framework outperforms the conventional schemes in both tasks, especially in the low SNR regime of interest. In addition, our framework is communication efficient, as our model transmits up to $89 \%$ less data than the baseline models. The problem of sharing the experience of the learned tasks to a new task via meta-learning is left for future research.

\appendices 
%


\bibliographystyle{IEEEtran}
\bibliography{IEEEabrv,references}

\begin{thebibliography}{10}
\providecommand{\url}[1]{#1}
\csname url@samestyle\endcsname
\providecommand{\newblock}{\relax}
\providecommand{\bibinfo}[2]{#2}
\providecommand{\BIBentrySTDinterwordspacing}{\spaceskip=0pt\relax}
\providecommand{\BIBentryALTinterwordstretchfactor}{4}
\providecommand{\BIBentryALTinterwordspacing}{\spaceskip=\fontdimen2\font plus
\BIBentryALTinterwordstretchfactor\fontdimen3\font minus \fontdimen4\font\relax}
\providecommand{\BIBforeignlanguage}[2]{{%
\expandafter\ifx\csname l@#1\endcsname\relax
\typeout{** WARNING: IEEEtran.bst: No hyphenation pattern has been}%
\typeout{** loaded for the language `#1'. Using the pattern for}%
\typeout{** the default language instead.}%
\else
\language=\csname l@#1\endcsname
\fi
#2}}
\providecommand{\BIBdecl}{\relax}
\BIBdecl

\bibitem{gunduz2022transmitting}
D.~Gündüz, Z.~Qin, I.~E. Aguerri, H.~S. Dhillon, Z.~Yang, A.~Yener, K.~K. Wong, and C.-B. Chae, ``Beyond transmitting bits: Context, semantics, and task-oriented communications,'' \emph{IEEE Journal on Selected Areas in Communications}, vol.~41, no.~1, pp. 5--41, 2023.

\bibitem{SC4IoV-XuVTM2023}
W.~Xu, Y.~Zhang, F.~Wang, Z.~Qin, C.~Liu, and P.~Zhang, ``Semantic communication for the internet of vehicles: A multiuser cooperative approach,'' \emph{IEEE Vehicular Technology Magazine}, vol.~18, no.~1, pp. 100--109, 2023.

\bibitem{6GV2XNoorProcIEEE2022}
M.~Noor-A-Rahim, Z.~Liu, H.~Lee, M.~O. Khyam, J.~He, D.~Pesch, K.~Moessner, W.~Saad, and H.~V. Poor, ``{6G} for vehicle-to-everything ({V2X}) communications: Enabling technologies, challenges, and opportunities,'' \emph{Proceedings of the IEEE}, vol. 110, no.~6, pp. 712--734, 2022.

\bibitem{10013736}
P.~M. de~Sant~Ana, N.~Marchenko, B.~Soret, and P.~Popovski, ``Goal-oriented wireless communication for a remotely controlled autonomous guided vehicle,'' \emph{IEEE Wireless Communications Letters}, vol.~12, no.~4, pp. 605--609, 2023.

\bibitem{8436265}
V.~Vakkuri and P.~Abrahamsson, ``The key concepts of ethics of artificial intelligence,'' in \emph{2018 IEEE International Conference on Engineering, Technology and Innovation (ICE/ITMC)}, 2018, pp. 1--6.

\bibitem{6GLEOEdge-LaiVTM2023}
Z.~Lai, H.~Li, Q.~Wu, Q.~Ni, M.~Lv, J.~Li, J.~Wu, J.~Liu, and Y.~Li, ``{Futuristic 6G Pervasive On-Demand Services: Integrating Space Edge Computing With Terrestrial Networks},'' \emph{IEEE Vehicular Technology Magazine}, vol.~18, no.~1, pp. 80--90, 2023.

\bibitem{LEOUAVJiaIOTJ2021}
Z.~Jia, M.~Sheng, J.~Li, D.~Niyato, and Z.~Han, ``{LEO}-satellite-assisted {UAV}: Joint trajectory and data collection for internet of remote things in {6G} aerial access networks,'' \emph{IEEE Internet of Things Journal}, vol.~8, no.~12, pp. 9814--9826, 2021.

\bibitem{RoleDLYuBITS2022}
W.~Yu, F.~Sohrabi, and T.~Jiang, ``Role of deep learning in wireless communications,'' \emph{IEEE BITS the Information Theory Magazine}, vol.~2, no.~2, pp. 56--72, 2022.

\bibitem{zhang2022goaloriented}
C.~Zhang, H.~Zou, S.~Lasaulce, W.~Saad, M.~Kountouris, and M.~Bennis, ``Goal-oriented communications for the {IoT} and application to data compression,'' \emph{IEEE Internet of Things Magazine}, vol.~5, no.~4, pp. 58--63, 2022.

\bibitem{10049005}
A.~Deb~Raha, M.~Shirajum~Munir, A.~Adhikary, Y.~Qiao, S.-B. Park, and C.~Seon~Hong, ``An artificial intelligent-driven semantic communication framework for connected autonomous vehicular network,'' in \emph{2023 International Conference on Information Networking (ICOIN)}, 2023, pp. 352--357.

\bibitem{zhang2022wireless}
M.~Zhang, Y.~Li, Z.~Zhang, G.~Zhu, and C.~Zhong, ``Wireless image transmission with semantic and security awareness,'' \emph{IEEE Wireless Communications Letters}, vol.~12, no.~8, pp. 1389--1393, 2023.

\bibitem{TOCShiWC2023}
Y.~Shi, Y.~Zhou, D.~Wen, Y.~Wu, C.~Jiang, and K.~B. Letaief, ``Task-oriented communications for {6G}: Vision, principles, and technologies,'' \emph{IEEE Wireless Communications}, vol.~30, no.~3, pp. 78--85, 2023.

\bibitem{TOCIBShaoJSAC2022}
J.~Shao, Y.~Mao, and J.~Zhang, ``Learning task-oriented communication for edge inference: An information bottleneck approach,'' \emph{IEEE Journal on Selected Areas in Communications}, vol.~40, no.~1, pp. 197--211, 2022.

\bibitem{SCV2V-SuTVT2023}
J.~Su, Z.~Liu, Y.-a. Xie, K.~Ma, H.~Du, J.~Kang, and D.~Niyato, ``{Semantic Communication-Based Dynamic Resource Allocation in D2D Vehicular Networks},'' \emph{IEEE Transactions on Vehicular Technology}, vol.~72, no.~8, pp. 10\,784--10\,796, 2023.

\bibitem{Asad_MTC}
M.~A. Ullah, K.~Mikhaylov, and H.~Alves, ``Enabling {mMTC} in remote areas: {LoRaWAN} and {LEO} satellite integration for offshore wind farm monitoring,'' \emph{IEEE Transactions on Industrial Informatics}, vol.~18, no.~6, pp. 3744--3753, 2022.

\bibitem{balle2016density}
J.~Ball\'e, V.~Laparra, and E.~P. Simoncelli, ``Density modeling of images using a generalized normalization transformation,'' \emph{arXiv}, 2016.

\bibitem{9504554}
E.~Eldeeb, M.~Shehab, and H.~Alves, ``A learning-based fast uplink grant for massive {IoT} via support vector machines and long short-term memory,'' \emph{IEEE IoT Journal}, vol.~9, no.~5, pp. 3889--3898, 2022.

\bibitem{1284395}
Z.~Wang, A.~Bovik, H.~Sheikh, and E.~Simoncelli, ``Image quality assessment: from error visibility to structural similarity,'' \emph{IEEE Transactions on Image Processing}, vol.~13, no.~4, pp. 600--612, 2004.

\bibitem{TRAFFIC_SIGN_DATA}
``{Chinese Traffic Sign Database},'' \url{http://www.nlpr.ia.ac.cn/pal/trafficdata/recognition.html}, accessed: June 14, 2023.

\end{thebibliography}
\end{document}